\documentstyle[prl,aps,multicol,epsfig]{revtex}
\begin{document}
\draft
\widetext
\title{Magnetotransport in Cuprates: a Test of
the Spin Fluctuation Model
}
\author{Branko Stojkovi\'c$^{a}$ and David Pines$^{a,b}$}
\address{
$^{a}$Center for Nonlinear Studies, Los Alamos National Laboratory,
Los Alamos, NM, 87545\\
$^{b}$University of Illinois at Urbana-Champaign,   Loomis Laboratory
of Physics, 1110 W. Green,  Urbana, IL, 61801
}

\date{\today}
\maketitle
\widetext
\begin{abstract}
\leftskip 54.8pt
\rightskip 54.8pt
We report on a simple calculation of the magnetotransport in
cuprate superconductors, based on the nearly
antiferromagnetic Fermi liquid (spin fluctuation) model.
We find that the model explains all important features seen
experimentally: the violation of K\"ohler's rule, the close
relationship between the Hall angle and the magnetoresistance, the
temperature dependence of the first high field correction to MR
and the doping dependence of the low field MR data. In addition,
the estimated values of $\omega_{\rm c}\tau$, calculated
using parameters obtained from the NMR measurements, yield
values in close agreement with those found experimentally for 
overdoped and optimally doped cuprates.
\end{abstract}
\begin{multicols}{2}
\narrowtext
Magnetotransport measurements of the Hall
effect\cite{don} and the magnetoresistance (MR)\cite{ong,bo}, 
in the normal state of cuprates\cite{don-book}, 
reveal the full peculiarity of these remarkable systems
and pose a major challenge to candidate theories
of high temperature
superconductivity. Harris et al\cite{ong} 
find that for modest magnetic fields,
$B$, the relative shift in the longitudinal resistivity,      
$\Delta \rho_{\rm xx}/\rho_{\rm xx}$, is
proportional to  $\Theta_{\rm H}^2$, where $\Theta_{\rm H}$, 
the Hall angle, $\equiv \rho_{\rm xy}/\rho_{\rm xx}$; 
this implies
that Kohler's rule is violated. Tyler et al\cite{bo}, who study the relative
magnetoresistance in overdoped cuprates placed in strong fields,
$B\sim 40-60$ T, find a departure from the modest field, $B^2$ behavior,
\begin{equation}
{\Delta\rho_{\rm xx}\over \rho_{\rm xx}} = {\alpha B^2\over 1+
\beta^2 B^2}
\label{eq:bo-1}
\end{equation}
a departure which occurs at higher threshold fields for
optimally doped or underdoped than for the overdoped cuprates.
In addition, Tyler et al\cite{bo} find that at a given
temperature and field,   $\Delta\rho_{\rm xx}/\rho_{\rm xx}$
is little changed as one goes from
optimally doped to overdoped samples. A further challenge is explaining
the specific values obtained in Ref.\ \cite{bo} for the parameter, 
$\omega_{\rm c}\tau$, where $\omega_{\rm c}$
is the cyclotron frequency and  $\tau$
is the effective relaxation rate.

         Below we show that the spin fluctuation model of the
superconducting cuprates, in which the system is described as a nearly
antiferromagnetic Fermi liquid, provides a natural explanation of all the
above results. In this model planar quasiparticles interact through an
effective magnetic interaction which is proportional to the dynamical spin
susceptibility, which is generally taken to be\cite{MMP90}
\begin{equation}
V_{\rm eff}^{\rm NAFL}({\bf q},\omega)=g^2 \chi_{\bf q}(\omega)=\frac{g^2
\chi_{\bf Q}}{1+\xi^2({\bf q}-{\bf Q})^2
- i {\omega\over \omega_{\rm sf}}}\,.
\label{MMP}
\end{equation}
Here $\chi_{\bf Q}=\alpha_{\rm AF} \xi^2$, with $\alpha_{\rm AF}$
constant, $\xi$ is the
AF correlation length, $\omega_{\rm sf}$ is the  frequency of the low
energy relaxational mode characteristic of a system which is nearly
antiferromagnetic (AF), and $g$ is the coupling constant. Because
the interaction is strongly peaked near the commensurate AF wavevector,
${\bf Q}=(\pi,\pi)$, 
it produces a pronounced anisotropy in quasiparticle behavior
near the Fermi surface. {\em Hot} 
quasiparticles, located near those parts of the
FS which can be connected by ${\bf Q}$, 
are so strongly scattered into one another
that their lifetimes and other properties are dramatically  different from
the behavior characteristic of a normal Landau Fermi liquid,  while
{\em cold}
quasiparticles, which are typically located near the diagonals,   
$|k_{\rm x}|=|k_{\rm y}|$, are
scattered rather weakly and display near Landau Fermi liquid
behavior. Detailed calculations, reported in Ref.\ \cite{SP97} 
show that by taking
this anisotropy into account one can deduce the quite different behavior of
hot and cold quasiparticles from single crystal measurements of 
$\rho_{\rm xx}$   and  $\rho_{\rm xy}$;
one finds in this way that $\tau_{\rm hot}\ll \tau_{\rm cold}$. 
These calculations also made it possible to
explain quantitatively the doping and temperature dependence of  both the
longitudinal conductivity and the Hall conductivity, while Schmalian et
al\cite{sps97} 
find that the spin fluctuation model can explain the weak pseudogap
behavior found in ARPES experiments.

        We begin our consideration of MR phenomena with the Zener-Jones
solution of the Boltzmann equation\cite{ziman}, 
which can be used to obtain the
conductivity tensor to arbitrary order in $B$:
\begin{equation}
\sigma_{\mu\nu} = 2e^2\sum_{\bf k} v^{\mu}_{\bf k}(1+{D})^{-1} 
\left\{v^{\nu}_{\bf k}\tau_{\bf k}\left({\partial f_0\over \partial
    \epsilon_{\bf k}}\right) \right\}
\end{equation}
where ${D}=e{\bf v}\times {\bf B}\cdot\nabla_{\bf k}$ is 
a differential operator, ${\bf v}_{\bf k}\equiv \nabla \epsilon_{\bf
k}$ is
the Fermi velocity and $\tau_{\bf k}$ the quasiparticle lifetime at a
point ${\bf k}$ on the FS. One can expand $(1+{\bf D})^{-1}$ in 
terms of the applied field: assuming ${\bf B}$
perpendicular to the $x-y$ plane, and switching from the sum to
integrals over the energy $\epsilon$ and the tangential component of
the momentum ${k}$ one finally obtains:
\begin{equation}
\sigma^{(n)}_{\mu\nu} = e^2\int_{\rm FS} 
{d{k} \over v_{\rm f}}
v^{\mu}_{\rm k} (-{D})^n (v^{\nu}_{\rm k}\tau_{\rm k})
\label{eq:zjk}
\end{equation}
where ${D}=eB\,v_{\rm k}\tau_{\rm k}(\partial/\partial{k})$.

Consider now applied fields low enough that  $\omega_{\rm c}\tau_{\bf k}\ll 1$.
In this limit  $\Delta \rho_{\rm xx}/\rho_{\rm xx} \approx
-\Delta\sigma_{\rm xx} / \sigma_{\rm xx} - \Theta_{\rm H}^2$.
Because of symmetry considerations, the linear in $B$ term in
Eq.\ (\ref{eq:zjk}) vanishes for $\sigma_{\rm xx}$ and it
is immediately clear that the low field MR is proportional to $B^2$, in
agreement with experiment and as in normal metals\cite{ong,bo,ziman}.
In this limit, the above formula for $\Delta\sigma_{\rm xx}$ yields
\begin{equation}
\Delta \sigma_{\rm xx} \approx  -e^2\int_{\rm FS}  
{d{k}\over (2\pi)^2 } \omega_{\rm c}^2 
\tau^3_{\rm {k}} v_{\rm {k}} \propto  
<\omega_{\rm c}^2  \ell_{{\rm k}}^3>\label{eq:sigxx_ell}
\end{equation}
where we have introduced the mean free path, $\ell_{\rm k} = 
\tau_{\rm k} v_{\rm k}$, and neglected terms which contain 
$\partial \tau_{\rm {k}}/
\partial{k}$, whose contribution turns out to be small.

Clearly, $\Delta \sigma_{\rm xx}$ and hence the MR are  governed 
primarily by 
{\em cold} quasiparticles whose  relaxation times are  the {\em largest}.
To take this anisotropy into account, we use 
the paramet\-ri\-za\-tion of Ref.\ \cite{SP97}, 
appropriate to a close-to-circular FS, in which the mean free
path of a point near the FS can be characterized by an angle $\theta$:
\begin{equation}
\ell^{-1}_\theta = \ell^{-1}_{\rm cold} {1+a\cos 4\theta\over 1-a}
\end{equation}
where  $a=(1-r)/(1+r)$ is the anisotropy parameter 
($r\equiv \ell_{\rm hot}/\ell_{\rm cold}$). We find, after some 
algebra, that
\begin{equation}
-{\Delta\sigma_{\rm xx}\over \sigma_{\rm xx}} \approx
{\omega_{\rm c}^2\ell_{\rm cold}^2\over 8v_{\rm f}^2} 
(3+2r + 3 r^2).\label{eq:del_rho}
\end{equation}
In the large anisotropy limit, 
on making use of the results for $\rho_{\rm xx}$ and $\cot \Theta_H$
(see Ref.\ \cite{SP97}):
\begin{mathletters}
\begin{equation}
\cot\Theta_H\approx {2\over \omega_{\rm c}\tau_{\rm cold}}
\end{equation}
\begin{equation}
\rho_{\rm xx} \approx {m_{\rm cold}\over  ne^2\tau_{\rm cold}\sqrt{r}},
\end{equation}
\end{mathletters}
this yields the experimental result of
Harris et al\cite{ong},
\begin{equation}
-{\Delta\sigma_{\rm xx}\over \sigma_{\rm xx}} \approx
{3\over 8}\omega_{\rm c}^2\tau_{\rm cold}^2\approx {3\over 2} \Theta_{\rm H}^2
\end{equation}
and
\begin{equation}
{\Delta\rho_{\rm xx}\over \rho_{\rm xx}} \approx {1\over 8}
\omega_{\rm c}^2\tau_{\rm cold}^2 \approx {1\over 2}
\Theta_{\rm H}^2.
\label{eq:drho_estimate}
\end{equation}
Thus,  although  hot quasiparticles contribute to both 
$\sigma_{\rm xx}$, 
and $\Delta\sigma_{\rm xx}$, the ratio, 
$\Delta\sigma_{\rm xx}/\sigma_{\rm xx}$, is determined entirely
by the cold quasiparticles, in accord with the $r\rightarrow 0$ 
limit of Eq. (7).
Since  $\cot \Theta_{\rm H}\propto T^2$ in optimally doped
cuprates\cite{ong}, 
Eq.\ (\ref{eq:drho_estimate}) 
explains why Harris et al observed $\Delta\rho_{\rm xx}/\rho_{\rm xx}\propto
T^{-4}$ in YBa$_2$Cu$_3$O$_7$ \cite{ong}. Moreover, since 
$\sigma_{\rm xx}$ involves both cold and hot quasiparticles, plotting
$\Delta\rho_{\rm xx}/\rho_{\rm xx}$ as a function of $B^2/\rho_{\rm xx}^2$ will
not produce a universal curve, independent of temperature, and 
hence in the spin-fluctuation model of cuprates 
K\"ohler's rule is
violated, again in agreement with experiment\cite{ong}.

\begin{figure}
\centerline{
\epsfig{file=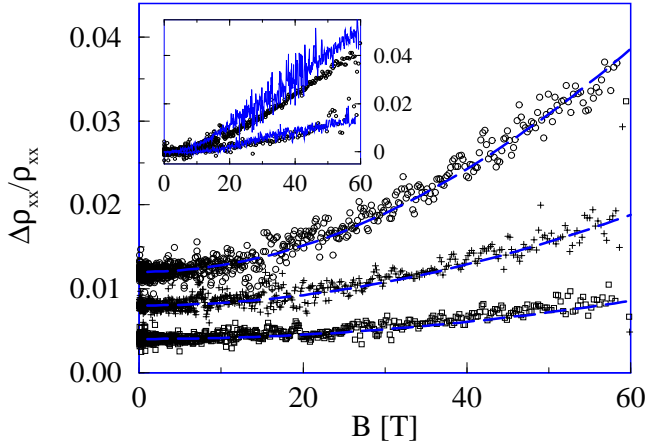,width=8.5cm, height=12cm, scale=0.85,
bb=050 275 350 700} }
\caption{The calculated relative (orbital)
magnetoresistance in an optimally doped Tl
2201 compound, is compared with
experiment (Ref.\ [3]): the symbols (lines), offset by 0.004 for
clarity,  correspond to 
the measured  (calculated) result at (top to bottom)
$T=140$, 200 and 260K.  The input parameters have been
obtained from NMR measurements (see Refs.\ [6,9,10]) and
are given in the text. Inset: the 
experimentally measured results of Tyler et al [3] for the 
low field orbital MR in the overdoped (solid lines) and optimally
doped (symbols)  Tl 2201  material at $T=120$ (top) and $T=180$ K (bottom). 
Note that the numerical values of the relative MR differ by less than
20\%, in agreement with Eq.\ (9).}
\label{fig:opt}
\end{figure}

Having established the important role played by the cold
quasiparticles, we consider next the magnitude of $\tau_{\rm cold}$. 
We use the expression derived in Ref.\ \cite{SP97}:
\begin{equation}
{\hbar\over \tau_{\rm cold}}={\hbar \over \tau_i} + \gamma 
k_B T{T\over T_0+T},\label{eq:tau_cold}
\end{equation}
where $\gamma=\alpha g^2 / 8\hbar v_{\rm f} (\Delta {k})$,
$T_0 =\omega_{\rm sf}\xi^2(\Delta {k})^2/\pi$, and
we introduce the constant scattering rate, $1/\tau_i$, 
independent of momentum and temperature, to model the
influence of static disorder.
We take the parameters which determine $\tau_{\rm cold}$ from our
earlier calculations of the resistivity and Hall angle for Tl 2201
(see Refs.\ \cite{SP97,CPS,TlNMR}): thus we assume that the spin 
fluctuations are in
the mean field regime, with  $\omega_{\rm sf} \xi^2 = {\rm const}=50$
meV, take  $\alpha=15$ states/eV,
$\hbar v_{\rm f}=0.5$ eV, $\Delta k\sim 1$ inverse lattice spacings,
$g=0.5$ eV and $m_*\approx 2m_e$, and so find $\gamma = 0.93$.
In addition, on assuming that the role played by static disorder
does not vary appreciably with doping, we estimate  (see below)
$\hbar/\tau_i=0.9$ meV from the resistivity measurements on an
overdoped sample.
Our calculated results [using Eqs.\ (\ref{eq:drho_estimate}) and
(\ref{eq:tau_cold})] for $\Delta\rho_{\rm xx} / \rho_{\rm xx}$ 
are compared with experiment in Fig.\ \ref{fig:opt}.
As may be seen, quantitative agreement with the experimental
results of Ref.\ \cite{bo} is found.
We remark that the approach we followed is self-consistent,
because for this choice of parameters $\omega_{\rm c} \tau_{\rm cold} = 0.5$
at $T\sim 200$ K and $B=60$ T.

We consider next the strong field behavior of the MR, i.e., what happens
when for cold quasiparticles the weak field condition,
$\omega_{\rm c} \tau_{\rm cold}\ll 1$
is no longer satisfied. For the above choice of parameters,
it is clear from Eq.\
(\ref{eq:tau_cold}) that as the temperature is lowered to 
$T\sim 50$ K this condition is violated.
 To study this we return to Eqs.\ (1) and (4) and consider
the contributions of order $B^4$. 
A straightforward, but lengthy calculation shows that in the limit of 
large relaxation rate anisotropy, $r \ll 1$, one finds
\begin{equation}
\beta^2 B^2 \approx (5/16)\omega_{\rm c}\tau_{\rm cold}.
\label{eq:beta}
\end{equation}
Of course, just as in the case of the parameter $\alpha$,
the exact value of $\beta$ depends, through $\tau_{\rm cold}$, 
on details  of band structure, although we do not expect it to change
dramatically as one moves 
from optimally doped to overdoped samples  within the same compound.
We emphasize that our estimate of $\beta$
is not the same as that obtained in the high field limit, where
$\omega_{\rm c}\tau_k\gg 1$ for all $k$. In this limit,
$\beta^2\propto <m^*/\tau>^{-1}$, and since the average scattering 
{\em rate} is dominated by the {\em hot} quasiparticles\cite{SP97},
$\beta$ would have a very different temperature dependence than that
given by Eq.\ (\ref{eq:beta}).
Experiment shows that
the value of $\omega_{\rm c}\tau$,
is comparatively low (e.g.,  in Tl 2201 at $T=40$ K and $B=60$ T
$\omega_{\rm c}\tau\sim 0.9$\cite{bo})
and therefore our estimate of $\beta$ is a more reasonable one.

Before comparing our strong field 
result to experiment it is important to take into account
the role of static disorder  
(the term $1/\tau_i$ in Eq.\ (\ref{eq:tau_cold}))
and  changes in the spin fluctuation and band
parameters as one goes from the optimally doped to the
overdoped material. 
 Using the experimental result for $\rho_{\rm xx}$\cite{bo}, which shows
a well defined residual resistivity, $\rho_0$, and comparing
$\rho_{\rm xx}(T)-\rho_0$ with $\rho_{\rm xx}(T)$, 
assuming that the temperature dependence of $\rho_{xx}$
comes only from the spin fluctuations, we estimate $\hbar/\tau_{\rm
i}=0.9$ meV, the value used above for the optimally doped
case. This estimate does not depend on the 
carrier density and  effective mass.
ARPES measurements show that for large doping
levels the observed FS agrees with Luttinger's theorem, so that
one expects  a somewhat larger value of $\Delta k$ in
the overdoped material, while 
the overall coupling constant may be slightly reduced\cite{CPS}.
These changes lead to a reduced  value
of $\gamma$ (of Eq.\ (\ref{eq:tau_cold})) as the doping level
increases.
Nevertheless, since the relative MR is governed by the
cold quasiparticles, which are not strongly scattered by
the spin fluctuations, we expect this change to be
slight and we take  $\gamma_{\rm over}\sim 0.9 \gamma_{\rm opt}$.

\begin{figure}
\centerline{\epsfig{file=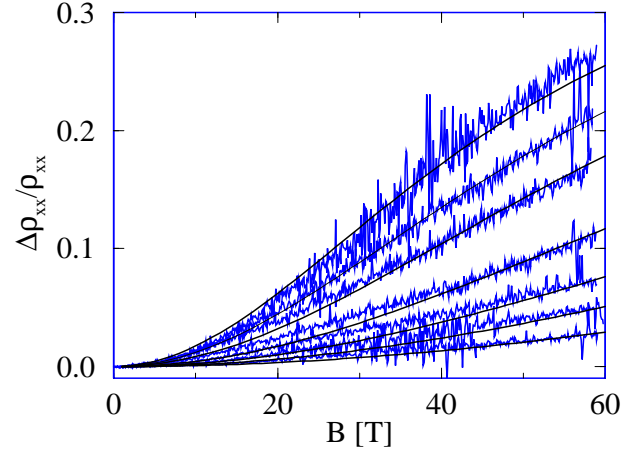,width=8.5cm, height=7cm, scale=0.85,
bb=050 450 350 700} }
\vskip5cm
\caption{The calculated relative magnetoresistance in an overdoped Tl 2201
compound, is compared with
experiment (Ref.\ [3]): the  results correspond to 
(top to bottom) $T=40$, 50,
60, 80, 100, 120 and 150K. The theoretical curve (solid lines)
is given by Eq.\ (1)
where the coefficients $\alpha$ and $\beta$ have been
determined from the spin fluctuation model, using parameters (see text)
obtained from NMR measurements in the optimally doped system (Refs. [6,9,10]).
Note the onset of high field effects at lower temperatures.}
\label{fig:over}
\end{figure}

Figure \ref{fig:over} compares our calculated result with
experiment \cite{bo}: we have made use of Eq.\ (\ref{eq:bo-1})
and assumed that  the
temperature dependence of $\alpha$ and $\beta$ comes 
directly from the temperature dependence of $\tau_{\rm k}$
[see Eqs.\ (\ref{eq:drho_estimate}) and (\ref{eq:beta})],
calculated using the above input parameters.
Again, reasonable quantitative agreement is found at all temperatures.

In arriving at the above results, we have assumed that $\gamma$ and $T_0$
and hence
$\tau_{\rm cold}$ do not vary appreciably as one goes from optimally
doped to overdoped systems. This means that
at optimal doping and higher doping levels, the {\em low field} relative MR, 
$\Delta\rho_{\rm xx}/\rho_{\rm xx}\propto (\omega_{\rm c}\tau_{\rm cold})^2$
must be weakly doping dependent as well. 
In the inset of Fig.\ \ref{fig:opt}  we demonstrate
that this is the case for Tl 2201: the results are within 20\% of each
other at all fields  and temperatures 
even though the superconducting transition
temperatures are vastly different ($T_{\rm c}=30$ and 80 K 
respectively). Moreover, the 
temperature dependence of the low field MR
in the two samples is essentially the 
same, even though the resistivity is qualitatively
different. The slightly larger 
value of the relative MR in the overdoped sample 
can be attributed to the role of the hot spots, which are
better
defined in overdoped materials, and hence can contribute to transport
more than they do in the optimally doped ones, as discussed below.

The onset of high field behavior is also seen 
in the Hall effect\cite{bo}: while the low field Hall resistivity is
given by $\rho_{\rm xy}=R_{\rm H} B$, where $R_{\rm H}$ is the Hall
constant, higher order terms lead to considerable
deviation of $\rho_{\rm xy}$ from linear in $B$ behavior\cite{bo}.
On using the same approximations as for the MR,
we have found that, in Tl 2201 at $T=40$ K and $B=60$ T,
the relative departure of $\rho_{\rm xy}$
from linearity, $1-R_{\rm H} B/\rho_{\rm xy}\approx 0.3$ in 
reasonable agreement with the experimental value of $\sim$0.25.

The quantitative agreement with experiment we find 
is remarkably good, given
the at first sight crude approximations we have made. 
For example, 
in Eq.\ (\ref{eq:drho_estimate}) we have 
neglected  terms which involve derivatives of $\tau_{\bf k}$ with 
respect to the momentum component parallel to the FS, while our
analysis assumes  that the FS in Tl 2201 is close to circular.
There are several reasons why these assumptions 
are  reasonable. 
First, in a NAFL the momentum dependence of
the relaxation rates can typically be factorized from their temperature
dependence (see Ref.\ \cite{SP97}). 
Therefore the inclusion of the terms involving these derivatives 
would only lead to
somewhat different momentum weights for the different temperature
dependencies seen in $\tau_{\rm k}$, 
but not to an overall changed behavior.
Second, the hot spots are typically not special symmetry points on
the FS. Hence contribution from the terms with derivatives
is substantially reduced by 
geometric factors. Still, inclusion of the remaining 
terms could, in principle, lead to rather different values of the parameter
$\alpha\propto \Delta\rho_{xx}/\rho_{xx}\omega_{\rm c}^2$.
For example, if the FS is close to rectangular, the value of 
$\alpha$ can be much larger than that quoted here.

We consider next the role played by  the hot quasiparticles, whose relaxation
rate is given by\cite{SP97}:
\begin{equation}
{1\over\tau_k} = {\alpha g^2 T \xi\over 4 v_{\rm f}} \left[
1 - 
(1+{\pi T\over \omega_{\rm sf}})^{-{1\over 2}}
\right].
\label{eq:tau_hot}
\end{equation}
In Ref.\ \cite{SP97} we have shown that only in the limit 
$T\gg\omega_{\rm sf}/\pi$, is
the behavior of the hot quasiparticles  anomalous compared
to the Fermi liquid behavior, leading to $r\rightarrow 0$.
The overdoped materials tend to have larger values of
$\omega_{\rm sf}$ (see Ref.\ \cite{BP95}), 
which, in turn, depends only moderately
on temperature. Both experiment and a comparison of Eqs.\
(\ref{eq:tau_cold}) and (\ref{eq:tau_hot}) shows that for
overdoped systems at low temperatures ($T\sim 50$ K) 
one no longer has  $r\ll 1$, so that higher order (in $r$)
terms must be taken into account. It is easy to show that these terms
lead to a reduction of $\alpha$ and in the limit $r\rightarrow 1$ (no
anisotropy) yield a vanishing MR, as expected for a  uniform scattering
rate\cite{ziman}.  Therefore, only
a full numerical calculation can confirm the quantitative agreement
with experiment seen here\cite{SP-MR}.

In conclusion, we have performed a simple analytical calculation
of the MR in cuprate superconductors, based on the spin fluctuation
(NAFL) model of cuprates. 
Quantitatively, the experimentally measured MR  appears to 
be in reasonable agreement 
with the magnetic measurements in cuprates, indicating
the close connection between the magnetic properties 
and the transport and therefore the importance of spin fluctuations
for the normal state of cuprates.
We find that the NAFL model gives a consistent description of all
experimental data to date. It naturally
explains  a close relationship between the orbital
MR and the Hall effect results\cite{ong}. It also predicts that the
experimentally observed deviations from the low field
behavior of the MR is due to onset, rather than the truly
high field behavior.

We are indebted to the authors of Ref.\ \cite{bo}  for communicating
their data prior to publication. We thank
G.\ Boebinger, F. Balakirev and A. Mackenzie 
for valuable discussions. This research has been
supported by the U.\ S.\ Department of Energy and by 
NSF through grant NSF-DMR 91-20000 
(Science and Technology Center for Superconductivity).

\vfill\eject

\end{multicols}

\end{document}